\documentclass[12pt]{article}
\usepackage{epsfig}

\def\3{\ss}

\def\qq{$Q^{2}$}
\def\g2{GeV$^{2}$}
\def\ds{D^{\ast}}
\def\d0{D^{0}}
\def\dspm{{\ds}^{\pm}}
\def\dskpi{ {\ds}^{+}~\rightarrow~\d0~\pi^{+}_{S}%
        \rightarrow~(K^{-}~\pi^{+})~\pi^{+}_{S} }
\def\dsk3pi{ {\ds}^{+}~\rightarrow~\d0~\pi^{+}_{S}%
        \rightarrow~(K^{-}~\pi^{+}~\pi^{+}~\pi^{-})~\pi^{+}_{S} }
\def\lumi{$36.9\pm~0.5$~pb$^{-1}$}
\def\bethe{$e^{+}p \rightarrow e^{+}{\gamma}p$}
\def\wrang{$130<W<280$\,GeV}

\def\ptrang{$p_{\perp}^{\ds}>2$\,GeV}
\def\ptrangold{$p_{\perp}^{\ds}>3$\,GeV}
\def\etarang{$-1.5<\eta^{\ds}<1.5$}

\def\qqrang{\qq~$<$~1\,\g2}
\def\totala{$3702\pm~136 $}
\def\totalb{$1397\pm~108$}

\def\xseca{$18.9\pm~1.2~({\it stat.})^{+1.8}_{-0.8}~
            ({\it syst.})$\,nb}
\def\xsecc{$2.57\pm~0.14~({\it stat.})^{+0.13}_{-0.08}~
            ({\it syst.})~
                                      ^{+0.29}_{-0.23}~
            ({\it energy~scale})$\,nb}
\def\xsecd{$1.65\pm 0.12~({\it stat.})^{+0.11}_{-0.06}~
            ({\it syst.})~
                                      ^{+0.20}_{-0.16}~
            ({\it energy~scale})$\,nb}

\def\xgo{x_\gamma^{\rm OBS}}

\def\ETJT{E_T^{jet}}
\def\ETJ1{E_T^{jet1}}
\def\ETJ2{E_T^{jet2}}
\def\ETJ#1{E_T^{jet#1}}

\def\ETAJ{\eta^{jet}}

\def\tablesup{\doublerulesep .5pt
  \def\0{\phantom{0}}\def\+{\phantom{-}}
  \def\PM##1##2{\mathop{^{+##1}_{-##2}}}
  \def\E##1##2##3{\pm ##1 \PM{##2}{##3}}
  \def\NL{\\[.5ex]}
  \def\RL{\hline\noalign{\vskip .5ex}}
  \def\RRL{\hline\hline\noalign{\vskip .5ex}}
  \centering
}
\begin{document}
\newcommand{\YJB}{$y_{\mathsf{JB}}$}

\title {{\normalsize\hfill\texttt{DESY~98-085}\\[3\bigskipamount]}
  \hspace*{-0mm}\large\rm \LARGE Measurement of inclusive $\dspm$ and
  associated dijet cross sections in photoproduction at HERA }
\author{ZEUS Collaboration} 
\date{ }

\maketitle

\begin{abstract}
\noindent
Inclusive photoproduction of $\dspm$ mesons has been measured for
photon-proton centre-of-mass energies in the range \wrang\ and photon
virtuality $Q^2 <$ 1\,\g2.  The data sample used corresponds to an
integrated luminosity of 37 pb$^{-1}$.  Total and differential cross
sections as functions of the $D^*$ transverse momentum and
pseudorapidity are presented in restricted kinematical regions and the
data are compared with next-to-leading order (NLO) perturbative QCD
calculations using the ``massive charm" and ``massless charm" schemes.
The measured cross sections are generally above the NLO calculations,
in particular in the forward (proton) direction.  The large data
sample also allows the study of dijet production associated with
charm.  A significant resolved as well as a direct photon component
contribute to the cross section.  Leading order QCD Monte Carlo
calculations indicate that the resolved contribution arises from a
significant charm component in the photon. A massive charm NLO parton
level calculation yields lower cross sections compared to the measured
results in a kinematic region where the resolved photon contribution
is significant.
\end{abstract}

\pagestyle{plain}

\thispagestyle{empty}

\vspace{3 cm}
\clearpage\begingroup
%===================================================================
%
%  MEMBER NAME  AUTH63 (ZEUS)     M  TEX
%
%  JH.: transformed to a format, which is suited as input for
%       CONVERT, which automatically creates author-indices
%
%  Don't remove lines starting with a percent sign %,
%  CONVERT may need them urgently !
%
%=====================================================================
\topmargin-1.cm
\evensidemargin-0.3cm
\oddsidemargin-0.3cm
\textwidth 16.cm
\textheight 680pt
\parindent0.cm
\parskip0.3cm plus0.05cm minus0.05cm
\def\3{\ss}
\pagenumbering{Roman}
                                    % this "%"s are for cosmetics only
                                                   %
\begin{center}
{                      \Large  The ZEUS Collaboration              }
\end{center}
  J.~Breitweg,
  M.~Derrick,
  D.~Krakauer,
  S.~Magill,
  D.~Mikunas,
  B.~Musgrave,
  J.~Repond,
  R.~Stanek,
  R.L.~Talaga,
  R.~Yoshida,
  H.~Zhang  \\
 {\it Argonne National Laboratory, Argonne, IL, USA}~$^{p}$
% ...
\newpage
$^{\    1}$ also at IROE Florence, Italy \\
% ...
$^{  19}$ now at ATLAS Collaboration, Univ. of Munich \\
% ...
$^{ 30}$ partially supported by the Foundation for German-Russian
Collaboration DFG-RFBR \\ 
\hspace*{3.5mm} (grant no. 436 RUS 113/248/3 and no. 436 RUS 113/248/2)\\
% ...
                                                           %
                                                           %
% \par         % if index listing & table fit to 1 page, put gap here
\newpage   % alternatively: go to newpage, if page is too small
                                                           %
% \institute_references_start    % do not touch or move this line !
                                                           %
\begin{tabular}[h]{rp{14cm}}
$^{a}$ &  supported by the Natural Sciences and Engineering Research
          Council of Canada (NSERC)  \\
$^{b}$ &  supported by the FCAR of Qu\'ebec, Canada  \\
$^{c}$ &  supported by the German Federal Ministry for Education and
          Science, Research and Technology (BMBF), under contract
          numbers 057BN19P, 057FR19P, 057HH19P, 057HH29P \\
$^{d}$ &  supported by the MINERVA Gesellschaft f\"ur Forschung GmbH,
          the German Israeli Foundation, the U.S.-Israel Binational
          Science Foundation, and by the Israel Ministry of Science \\
$^{e}$ &  supported by the German-Israeli Foundation, the Israel Science
          Foundation, the U.S.-Israel Binational Science Foundation, and by
          the Israel Ministry of Science \\
$^{f}$ &  supported by the Italian National Institute for Nuclear Physics
          (INFN) \\
$^{g}$ &  supported by the Japanese Ministry of Education, Science and
          Culture (the Monbusho) and its grants for Scientific Research \\
$^{h}$ &  supported by the Korean Ministry of Education and Korea Science
          and Engineering Foundation  \\
$^{i}$ &  supported by the Netherlands Foundation for Research on
          Matter (FOM) \\
$^{j}$ &  supported by the Polish State Committee for Scientific
          Research, grant No.~115/E-343/SPUB/P03/002/97, 2P03B10512,
          2P03B10612, 2P03B14212, 2P03B10412 \\
$^{k}$ &  supported by the Polish State Committee for Scientific
          Research (grant No. 2P03B08614) and Foundation for
          Polish-German Collaboration  \\
$^{l}$ &  partially supported by the German Federal Ministry for
          Education and Science, Research and Technology (BMBF)  \\
$^{m}$ &  supported by the Fund for Fundamental Research of Russian Ministry
          for Science and Edu\-cation and by the German Federal Ministry for
          Education and Science, Research and Technology (BMBF) \\
$^{n}$ &  supported by the Spanish Ministry of Education
          and Science through funds provided by CICYT \\
$^{o}$ &  supported by the Particle Physics and
          Astronomy Research Council \\
$^{p}$ &  supported by the US Department of Energy \\
$^{q}$ &  supported by the US National Science Foundation \\
\end{tabular}
                                                           %
% \institute_references_end     % do not touch or move this line !

\clearpage                                                  %
\endgroup
\pagenumbering{arabic}

\section{Introduction}

In photoproduction processes at HERA, a quasi-real photon ($Q^2\simeq
0$) is emitted by the incoming electron or positron, and interacts
with the proton.  Such a photon has a hadronic component, which can be
assigned a partonic structure. At leading order (LO) in QCD, two types
of process take part in photoproduction: direct photon processes,
where the photon couples as a point-like particle to a parton from the
proton, and resolved photon processes, where one of the partons in the
photon scatters on a parton in the proton.  The light quark structure
of the photon has been extensively studied in photon-photon collisions
at $e^+ e^-$ storage rings~\cite{soldner}, whilst there is little
information at present on the charm content of the photon.  HERA jet
studies have shown some sensitivity to the gluon content of the
photon~\cite{H1, dijet}, which is still poorly known.  In this paper
we present a study of charm photoproduction. Here, the direct process
is photon-gluon fusion, $\gamma g\to c\bar c$, while charm quarks in
the parton distributions of the photon and the proton can lead to
processes of the type $cg\to cg$, known as charm flavour excitation.

The photoproduction of heavy quarks such as charm can be calculated
using perturbative QCD (pQCD) with a hard scale given by the heavy
quark mass or by the high transverse momentum of the produced partons.
Two types of QCD NLO calculations are available for comparison with
measurements of charm photoproduction. The massive charm
approach~\cite{NLOdiff} assumes that gluons and light quarks (u,d,s)
are the only active partons within the proton and the photon, so that
charm is only produced dynamically in the hard process. In the
massless charm approach~\cite{kniehl, kniehl2, cacciari} charm is
treated as an additional active flavour.  The massive approach is
expected to be superior for $p_{\perp}^2\simeq m_c^2$ while the
massless one is expected to describe the data better for $p_{\perp}^2
\gg m_c^2$~\cite{Forshaw}, where $p_{\perp}$ and $m_c$ are the
transverse momentum and mass of the charm quark.  In NLO calculations,
direct and resolved components cannot be unambiguously separated.  The
massless charm calculations take into account charm excitation
processes and thus predict, for a given factorisation scale, a larger
resolved component in comparison with the massive calculation.
Therefore, it is interesting to compare the predictions of these
models to data and to investigate the sensitivity of the experimental
results to the partonic content of the photon and specifically to the
charm excitation contribution.

In the analysis described in this paper, charm was tagged by
identifying $\dspm$(2010) mesons in the final state via the charged
products of their decay.  $\ds$ mesons are reconstructed through the
two decay modes\footnote{In this analysis $\dspm$(2010) are referred
  to as $\ds$ and the charge conjugated processes are also included.}:
\begin{equation}
   \dskpi,
   \label{kpi}
\end{equation}
\begin{equation}
   \dsk3pi.
   \label{k3pi}
\end{equation}

\noindent
The small mass difference $\Delta M = M(\ds) - M(\d0) = 145.42 \pm
0.05 $\,MeV~\cite{PDG} yields a low momentum pion (``soft pion'',
$\pi_{S}$) from the $D^*$ decay and prominent signals just above the
threshold of the $ M(K\pi\pi_{S}) - M(K\pi) $ and
$M(K\pi\pi\pi\pi_{S}) - M(K\pi\pi\pi) $ distributions, where the phase
space contribution is highly suppressed~\cite{DELTAM}.

We present measurements of integrated and differential cross sections
for $\ds$ mesons produced in restricted kinematic regions in
$p_{\perp}^{\ds}$ and $\eta^{\ds}$. Here $\eta^{\ds}$ is the
pseudorapidity of the $\ds$, defined as $-\ln(\tan(\theta /2))$, where
the polar angle $\theta$ is taken with respect to the proton beam
direction.\footnote{We use the standard ZEUS right-handed coordinate
  system, in which $X = Y = Z = 0$ is the nominal interaction point
  and the positive $Z$-axis points in the direction of the proton beam
  (referred to as the forward direction).}  The data sample is larger
by more than an order of magnitude compared to our previous
study~\cite{dstar94}, which allows an accurate measurement of the
differential cross sections in both $p_{\perp}^{\ds}$ and $\eta^{\ds}$
and thus a more stringent test of the NLO QCD predictions.

The improved statistics of the $D^*$ sample allows, for the first
time, the study of dijet photoproduction in association with charm.
In such events, the fraction $\xgo$ of the photon momentum which
participates in the dijet production can be measured~\cite{dijet}.
This quantity is sensitive to the relative contributions of resolved
and direct processes~\cite{DIRECT}.  In LO QCD direct photon events at
the parton level have $\xgo$=1, while resolved photon events populate
low values of $\xgo$.

\section{Experimental Conditions}

The data presented in this analysis were collected with the ZEUS
detector at HERA during the 1996 and 1997 running periods, where a
positron beam with energy $E_e$=27.5~GeV collided with a proton beam
with energy $E_p$=820~GeV.  The data sample corresponds to an
integrated luminosity of \lumi .  A detailed description of the ZEUS
detector can be found in refs.~\cite{ZEUS2,ZEUS1}.  Here we present a
brief description of the components relevant to the present analysis.

Charged particles are measured by the Central Tracking Detector
(CTD)~\cite{CTD} which operates in a magnetic field of 1.43\,T
provided by a thin superconducting solenoid. The CTD is a drift
chamber consisting of 72~cylindrical layers, arranged in 9 superlayers
covering the polar angle region $15^\circ < \theta < 164^\circ$.  The
transverse momentum resolution for full length tracks is
$(\sigma_{p_{\perp}}/p_{\perp})^{track}= 0.005
p_{\perp}^{track}\bigoplus 0.016 $ ($p_{\perp}^{track}$ in~GeV).  The
CTD was also used to establish an interaction vertex for each event.

Surrounding the solenoid is the uranium-scintillator sampling
calorimeter (CAL)~\cite{CAL}.  The CAL is hermetic and consists of
5918 cells each read out by two photomultiplier tubes. Under test beam
conditions, the CAL has an energy resolution of $0.18/\sqrt{E}$ for
electrons and $0.35/\sqrt{E}$ for hadrons ($E$ in~GeV).  The effects
of uranium noise were minimised by discarding cells in the inner
(electromagnetic) or outer (hadronic) sections if they had energy
deposits of less than 60\,(110)\,MeV. For cells without energy
deposits in neighbouring cells this cut was increased to
80\,(140)\,MeV.

The luminosity was measured from the rate of the bremsstrahlung
process \bethe, where the photon is measured by a
calorimeter~\cite{LUMI} located at $Z = -107$\,m in the HERA tunnel.

The ZEUS detector uses a three level trigger system~\cite{ZEUS1}.  At
the first level trigger (FLT) the calorimeter cells were combined to
define regional and global sums which were required to exceed various
CAL energy thresholds.  In addition, at least one CTD track coming
from the $ep$ interaction region was required.

At the second level trigger, beam-gas events were rejected by
exploiting the excellent timing resolution of the calorimeter and by
cutting on the quantity $\Sigma_{i}(E-p_{Z})_{i} > 8$~GeV, where the
sum runs over all calorimeter cells and $p_{Z}$ is the $Z$ component
of the momentum vector assigned to each cell of energy $E$.  In
addition, events were rejected if the vertex determined by the CTD was
not compatible with the nominal $ep$ interaction point.

At the third level trigger (TLT) the full event information was
available.  Calorimeter timing cuts were tightened in order to reject
the remaining beam-gas events.  At least one combination of tracks
detected in the CTD was required to be within wide mass windows around
the nominal values in $\Delta M$ and in $M(K\pi)$ ($M(K\pi\pi\pi)$)
for reaction~(1) (reaction~(2)).  In addition, cuts were made on the
transverse momenta of tracks associated with these $D^*$ candidates
and $p_{\perp}^{\ds}$ was required to be above 1.8~GeV for
reaction~(1) and above 3.3~GeV for reaction~(2).  For the measurement
of $D^*$ in association with jets, an alternative trigger strategy is
possible at the TLT, based upon the jets themselves.  The jet
reconstruction algorithm used the CAL cell energies and positions to
identify jets. Events were required to have at least two jets, each of
which has a transverse energy $E_{T,cal}^{jet} > 4$~GeV and
pseudorapidity $\eta^{jet} < 2.5$. This strategy is used as a
cross-check for the results on dijets in association with charm.

\section{Analysis}

\subsection{Offline Data Selection}

The event sample was processed using the standard offline ZEUS
detector calibration and event reconstruction code.  To define an
inclusive photoproduction sample, the following requirements were
imposed:
\begin{itemize}
\item{} A reconstructed vertex with at least three associated tracks.
\item{} No scattered positron found in the CAL by the algorithm
  described in ref.~\cite{DIRECT}.  This requirement removes neutral
  current deep inelastic scattering (DIS) events, thereby restricting
  \qq~to below $\simeq$ 1\,GeV$^2$.  The corresponding median \qq~in
  our photoproduction sample is estimated from Monte Carlo (MC)
  simulations to be $\simeq~3{\cdot}10^{-4}$\,\g2.
\item{} $115 < W_{JB} < 250$~GeV, where $W_{JB} =
  \sqrt{4y_{JB}E_{p}E_{e}} $. Here $W_{JB}$ and \linebreak
  $y_{JB}={\Sigma_{i}(E-p_{Z})_{i}/2E_{e}}$ are the
  Jacquet-Blondel~\cite{JB} estimators of $W$ and $y$, respectively,
  and $y$ is the fraction of the positron beam energy taken by the
  photon.  The value of $W_{JB}$ was determined from the energy
  deposits in the uranium calorimeter.  The lower $W_{JB}$ cut rejects
  events from a region where the acceptance is small because of the
  trigger requirements. The upper cut rejects possible background from
  DIS events in which the scattered positron had not been recognised.
  A systematic shift in the reconstructed values of $W_{JB}$ with
  respect to the true $W$ of the event, due to energy losses in
  inactive material in front of the calorimeter and particles lost in
  the beam pipe, was corrected~\cite{DIRECT, paper_93}, using the MC
  simulation of the detector described in section~4.  The centre of
  mass energy range covered by the photoproduction sample is then
  \wrang, corresponding to $0.19 < y < 0.87$.
\end{itemize}

\subsection{ \bf \boldmath Reconstruction of $\ds$ Candidates }

A $\ds$ reconstruction algorithm was applied to all selected events.
It uses the mass difference technique to suppress the high background
due to random combinations from non-charm events, which have a much
higher cross section.  Only tracks associated with the event vertex
and having $p_{\perp}^{track} > 0.15$\,GeV and $|\eta^{track}| < 1.75
$ were included in the combinations.

Reconstructed tracks in each event were combined to form $\d0$
candidates assuming the decay channels~(\ref{kpi}) or (\ref{k3pi}).
For both cases, $\d0$ candidates were formed by calculating the
invariant mass $M(K\pi)$ or $M(K\pi\pi\pi)$ for combinations having a
total charge of zero. No particle identification was used, so kaon and
pion masses were assigned in turn to each particle in the combination.
Transverse momenta of $p_{\perp}^{track} > 0.5$\,GeV were required for
all tracks of channel~(\ref{kpi}) and for the track taken to be the
kaon for channel~(\ref{k3pi}). Pion candidates in the latter channel
were required to have $p_{\perp}^{track} > 0.3$\,GeV.  An additional
track, assumed to be the soft pion, $\pi_{S}$, with a charge opposite
to that of the particle taken as a kaon, was then added to the $\d0$
candidate. The mass difference $\Delta M = M(K\pi\pi_{S}) - M(K\pi)$
for channel~(\ref{kpi}) or $M(K\pi\pi\pi\pi_{S}) - M(K\pi\pi\pi)$ for
channel~(\ref{k3pi}) was evaluated. The reconstructed $\ds$ candidates
were required to be in the pseudorapidity range \etarang~, for which
the CTD acceptance is high.

To comply with the $p_{\perp}^{D^*}$ cut applied at the TLT, we
required $p_{\perp}^{K\pi\pi_{S}}>2$\,GeV for channel~(\ref{kpi}).
The number of decay particles in channel~(\ref{k3pi}) is larger: to
improve the signal to background ratio, we required
$p_{\perp}^{K\pi\pi\pi\pi_{S}}>4$\,GeV for this channel. Since more
combinatorial background exists in the forward direction as well as in
the region of low $p_{\perp}^{\ds}$, an additional cut,
$p_{\perp}^{\ds} / E_{\perp}^{\theta > 10^{\circ}} > 0.1$, was applied
to both channels.  Here $E_{\perp}^{\theta > 10^{\circ}}$ is the
transverse energy outside a cone of $\theta = 10^{\circ}$ defined with
respect to the proton direction.  This cut, as verified by MC studies,
removed a significant fraction of the background whilst preserving
$99\%$ of the $D^*$ signal.

The $\Delta M$ distributions of channel~(\ref{kpi}) and
channel~(\ref{k3pi}) for combinations with $M(K\pi)$ or
$M(K\pi\pi\pi)$ between 1.80 and 1.92~GeV are shown in Fig.1.  Clear
peaks at the nominal value of $M(\ds) - M(\d0)$ are evident.  MC
studies have shown that the contribution of other $\d0$ decay modes to
the $\Delta M$ peak is small and can be neglected.

The $\Delta M$ signals were fitted, using a maximum likelihood method,
to a sum of a Gaussian (describing the signal) and a functional form
(describing the background shape) of \mbox{$A\cdot~(\Delta
  M-m_{\pi})^{B}$}.  The mass values obtained were $\Delta M =
145.45\pm 0.02~({\it stat.})$\,MeV for channel~(\ref{kpi}) and
$145.42\pm 0.05~({\it stat.})$\,MeV for channel~(\ref{k3pi}), in
agreement with the PDG value~\cite{PDG}.  The width of the signals
were $\sigma = 0.68\pm 0.02$\,MeV and $\sigma = 0.72\pm 0.05$\,MeV,
respectively, in agreement with our MC simulation.

To determine the background under the peak for channel~(\ref{kpi}),
combinations in the same $M(K\pi)$ range, in which both tracks forming
the $\d0$ candidates have the same charge, with $\pi_S$ having the
opposite charge, were used. These are referred to as wrong charge
combinations.  The $\Delta M$ distribution from such combinations is
shown as the dashed histogram in Fig.\,1a.  The inset to Fig.\,1a
shows the $M(K\pi)$ distribution from combinations having a mass
difference in the range $143 < \Delta M < 148$~MeV. A $\d0$ peak is
clearly observed. The dashed histogram shows the wrong charge
combinations defined above. The excess of events with respect to the
wrong charge distribution below the $\d0$ region originates mostly
from $\d0$ decays involving neutral pions~\cite{dstar94}.  The number
of reconstructed $\ds$ mesons in channel~(\ref{kpi}) was determined by
subtracting the wrong charge distribution after normalising it to the
right charge distribution in the range $150 < \Delta M < 170$\,MeV.
After subtracting the background from the $\Delta M$ distribution of
Fig.\,1a, a signal of {\totala} $\ds$ events was obtained for
$p_{\perp}^{D^*} > 2$\,GeV.

Side band subtraction, close to the signal region, was used to
estimate the \linebreak background under the $\Delta M$ signal of
channel~(\ref{k3pi}).  The side bands, $1.70<M(K\pi\pi\pi)<1.80$~GeV
and $1.92<M(K\pi\pi\pi)<2.02$~GeV, were normalised to the region
$148<\Delta M<165$~MeV (dashed histogram in Fig.\,1b).  This
subtraction removed the combinatorial background coming from events or
tracks in which no $D^*$ decaying through this channel is produced,
and part of the background due to the mass misassignment of the kaon
and pion candidates with the same charge from the $\d0$ decay.  The
remaining background, coming from the mass misassignment, is
reproduced in the MC acceptance calculations.  The inset to Fig.\,1b
shows the $M(K\pi\pi\pi)$ distribution from combinations having a mass
difference in the range $143 < \Delta M < 148$~MeV. A $\d0$ peak is
clearly observed.  The total number of $\ds$ mesons in channel~(2)
extracted for $p_{\perp}^{D^*} > 4$\,GeV from the $\Delta M$
distribution with the side band subtraction was {\totalb}.

\subsection{Jet Reconstruction}

For the measurement of charmed dijet events, the KTCLUS cluster
algorithm~\cite{cluster} has been implemented in its ``inclusive"
mode~\cite{ellis}.  In this algorithm, jets are unambiguously defined
at the hadron, parton and CAL levels.  Using the $p_T$ recombination
scheme~\cite{ellis}, the parameters of the jets are calculated as:
$\ETJT=\sum_i E_{T_i}$; $\eta^{jet}=(1/\ETJT)(\sum_i E_{T_i}\eta_i)$;
\linebreak $\phi^{jet}=(1/\ETJT)(\sum_i E_{T_i}\phi_i)$.  The sums run
over all calorimeter cells, hadrons or partons belonging to the
corresponding jet.  Here $E_{T_i}$, $\eta_i$ and $\phi_i$ are the
transverse energy, pseudorapidity and azimuthal angle.

For the analysis of charm with associated dijets, events containing a
$D^*$ meson in channel~(1) with $p_{\perp}^{K\pi\pi_{S}} > 3$~GeV were
used. The events were also required to have at least two jets with
$|\ETAJ| < 2.4$ and a reconstructed $E_{T,cal}^{jet}>5$~GeV.  With
this selection, $587\pm 41$ events were found after subtraction of the
wrong charge background.  In addition, an analysis with
$E_{T,cal}^{jet} > 4$~GeV was performed, yielding $971\pm 52$ events.
The distribution of the distance between a $\ds$ candidate and the jet
closest to it in the $\eta^{jet}$- $\phi^{jet}$ space shows that the
measured $D^*$ belongs to one of the two jets.  In more than $80\%$ of
the cases, this distance was less than 0.2, which is consistent with
the observed hard fragmentation of heavy quarks~\cite{PETER}.

\section{Monte Carlo Simulation}

The MC programs PYTHIA 6.1~\cite{PYTHIA} and HERWIG 5.9~\cite{HERWIG}
were used to model the hadronic final states in charm production and
to study the efficiency of the cuts used in the data selection.  Both
programs are general purpose generators including a wide range of
photoproduction processes.

Large samples of charm events were generated for channels~(\ref{kpi})
and (\ref{k3pi}) using both MC programs. Direct and resolved photon
events, including charm excitation, were generated using as a
reference sample the MRSG~\cite{MRSG} parametrisation for the proton
and \linebreak GRV-~G~HO~\cite{GRV} for the photon.  These samples
have at least ten times the statistics of the data, so their
contribution to the statistical error is negligible.  To check the
sensitivity of the results to the choice of the structure function,
the reference samples were reweighted to simulate other parton
distributions of both the proton and the photon.  The MC studies
showed that, in the kinematic range used here, the results are
insensitive to contributions from charm excitation in the proton.

In order to include photoproduced $D^*$ mesons originating from $b$
quark events, a sample of such events was generated with a ratio to
the charm sample proportional to the cross section ratio of the two
processes used in the MC ($\simeq 1:100$).  Within the kinematic range
of the inclusive $D^*$ analysis, the contribution of $b$ quark
production to the $D^*$ cross section is estimated to be $\simeq 5\%$.
For the kinematic range of dijets in association with charm the
corresponding estimate is $\simeq 10\%$.

Events containing at least one $\ds$ decaying into channel~(\ref{kpi})
or (\ref{k3pi}) were processed through the standard ZEUS detector and
trigger simulation programs and through the same event reconstruction
package used for offline data processing.  Tracks were reconstructed
both in the TLT and the offline simulations.  The MC efficiency of the
tracking trigger was checked using the jet trigger described in
section~2 and found to be consistent with the data.  Satisfactory
agreement was observed between the CTD transverse momentum resolution
in the MC samples and the data.

An additional sample of events was generated using multiparton
interactions (MI) in HERWIG~\cite{butt} as an attempt to simulate the
energy from additional softer scatters (``underlying event").

\section{\boldmath Measurement of Inclusive $D^*$ Cross Sections}

The improved trigger and detector conditions compared to that used for
our previous results~\cite{dstar94} allow measurements of the
inclusive $ep\,\rightarrow\,\ds X$ cross sections in a wider kinematic
region: \ptrang\ and \etarang.  The integrated ${\ds}$ cross section
in the above region for \qqrang, \wrang~ was calculated using the
formula $ \sigma_{ep\rightarrow{D^*}X}={N_{corr}^{\ds}/{{\cal L} B}},$
where $N_{corr}^{\ds}$ is the acceptance-corrected number of $\ds$,
$B$ is the combined $\ds$ and $\d0$ decay branching ratios ($0.0262\pm
0.0010$ for channel~(1) and $0.051\pm 0.003$ for
channel~(2))~\cite{PDG} and ${\cal L}$= \lumi~is the integrated
luminosity.

In order to obtain $N_{corr}^{\ds}$, a correction factor $\omega_i$,
defined as the number of generated divided by the number of
reconstructed $\ds$ mesons, was calculated for channel~(1) from the MC
simulation using a three-dimensional grid in the quantities
$p_{\perp}^{\ds}$, $\eta^{\ds}$ and $W_{JB}$.  The index $i$
corresponds to a given grid bin.  All $\ds$ data candidates in a grid
bin were corrected by the appropriate $\omega_i$, yielding
$N^{\ds}_{corr}= \Sigma_{i} \omega_i (N^{\ds}_{rec})_i$.  Here
$(N^{\ds}_{rec})_i$ is the number of reconstructed $\ds$ candidates in
bin $i$.  For channel~(2) a one dimensional bin-by-bin unfolding
procedure was used.

The reference MC used to calculate the acceptance for channel~(1) was
HERWIG.  For channel~(2) PYTHIA was used, since HERWIG does not
reproduce the decay widths of resonances which contribute to the
$K\pi\pi\pi$ final state~\cite{PDG}.  Results obtained from the
alternative MC were used in each channel to estimate the systematic
uncertainties.

Table\,1 summarises the results for $N^{\ds}_{rec}$ after background
subtraction and the integrated cross sections for both decay channels
with various $p_{\perp}^{\ds}$ cuts.  The first error is statistical
and the second is the combined systematic uncertainty.  The overall
scale uncertainties ($\pm 1.4 \%$ from the luminosity measurement, and
$\pm3.7 \%$ or $\pm 5.7 \%$ from the branching ratios~\cite{PDG} of
channels~(1) or (2) respectively) were not included in the combined
systematic errors.

The differential cross sections $d\sigma/dp_{\perp}^{D^*}$ and
$d\sigma/d\eta^{D^*}$ were measured using the same procedure.  The
combinatorial background was subtracted bin-by-bin from each
distribution using the methods described above.  The
$d\sigma/dp_{\perp}^{D^*}$ distribution is shown in Fig.\,2 for $-1.5
< \eta^{D^*} < 1.5$ for channels~(\ref{kpi}) all and~(\ref{k3pi}) and
listed in Table~2 for channel (1).  The $d\sigma/d\eta^{D^*}$
distributions for $p_{\perp}^{D^*} > 2$ and 3\,GeV for
channel~(\ref{kpi}) are shown in Fig.\,3(a,b) and for $p_{\perp}^{D^*}
> 4$ and 6\,GeV for both channels in Fig.\,3(c,d).  In Table~3 the
$d\sigma/d\eta^{D^*}$ values are listed for channel~(1).

\begin{table}[t]\tablesup
  \tabcolsep 0pt
  \begin{tabular*}{\hsize}{@{\extracolsep{\fill}}lllll}\RRL
    & $p_{\perp}^{\ds}>2$~GeV & $p_{\perp}^{\ds}>3$~GeV & 
    $p_{\perp}^{\ds}>4$~GeV & $p_{\perp}^{\ds}>6$~GeV
    \NL\RL
    $N^{\ds}_{\rm rec}(K\pi\pi_{\rm s})$ &
    \totala &$ 2619\pm 82 $&$ 1505\pm 50 $&$ 410\pm 24 $
    \NL
    $N^{\ds}_{\rm rec}(K\pi\pi\pi\pi_{\rm s})$ &
    {}      &{}            & \totalb      &$ 411\pm 40 $\NL
    \RL
    $\sigma_{\rm data}(K\pi\pi_{\rm s})$ [nb]&
    $ 18.9\E{1.2}{1.8}{0.8}$ & $9.17\E{0.35}{0.40}{0.39}$ &
    $4.24\E{0.16}{0.16}{0.14}$ & $0.948  \E{0.061  }{0.046  }{0.047}$
    \NL
    $\sigma_{\rm data}(K\pi\pi\pi\pi_{\rm s})$ [nb]&&&
    $4.22\E{0.33}{0.41}{0.15}$ & $0.991  \E{0.098  }{0.099  }{0.063}$
    \NL
    $\sigma_{\rm massive}$ [nb]               
    & 13.1 & 5.43 & 2.46 & 0.665 \NL
    $\sigma_{\rm massless}$~\cite{kniehl} [nb]  
    & 25.3 & 8.50 & 3.37 & 0.739 \NL
    $\sigma_{\rm massless}$~\cite{cacciari} [nb] 
    & 17.4 & 5.83 & 2.34 & 0.520 \NL
    \RRL
  \end{tabular*}
  \caption{Number of reconstructed $D^*$ mesons after background
    subtraction and integrated cross sections, $\sigma_{ep\rightarrow\ds X}$,
    for $Q^2~<~1$~GeV$^2$, \wrang, $-1.5 < \eta^{D^*} < 1.5$ and various
    $p_{\perp}^{\ds}$ cuts. Predictions of the NLO QCD calculations
    are given for the reference parameters and parton density functions 
    (see section~6).  The first error is statistical and the second
    is systematic.  Overall normalisation uncertainties due to
    luminosity measurement ($\pm 1.4\%$) and to $\ds$ and $\d0$ decay
    branching ratios ($\pm3.7 \%$ for channel~(1) and $\pm 5.7 \%$ for
    channel~(2)) are not included in the systematic errors.}\smallskip
\end{table}

\begin{table}[p]\tablesup
  \begin{tabular}{cc}\RRL
    $p_\perp^{\ds}$ (range) GeV&
    $d\sigma/dp_\perp^{\ds} $ (nb/GeV)\NL\RL
    2.458 ( 2\0--\03 )&$ 9.68\0\E{1.16\0}{1.55\0}{0.56\0}$\NL
    3.464 ( 3\0--\04 )&$ 4.94\0\E{0.31\0}{0.30\0}{0.31\0}$\NL
    4.469 ( 4\0--\05 )&$ 2.22\0\E{0.13\0}{0.10\0}{0.11\0}$\NL
    5.470 ( 5\0--\06 )&$ 1.076 \E{0.073 }{0.071 }{0.043 }$\NL
    6.902 ( 6\0--\08 )&$ 0.328 \E{0.024 }{0.020 }{0.013 }$\NL
    9.672 ( 8\0--12  )&$ 0.067 \E{0.008 }{0.004 }{0.006 }$\NL
    \RRL
  \end{tabular}
  \caption{The differential cross section $d\sigma/dp_\perp^{\ds}$ for
    channel~(1) as function of $p_\perp^{\ds}$ for the kinematic
    region of Fig.2.  The $p_\perp^{\ds}$ points are given at the
    positions of the average values of an exponential fit in each
    bin. The $p_\perp^{\ds}$ range is given in brackets.  The first
    error is statistical and the second is systematic.  Overall
    normalisation uncertainties due to luminosity measurement ($\pm
    1.4\%$) and to $\ds$ and $\d0$ decay branching ratios ($\pm3.7
    \%$) are not included in the systematic errors.}\smallskip
\end{table}

\begin{table}[p]\tablesup
  \begin{tabular}{ccc}\RRL
    $\eta^{\ds}$ range&\multicolumn2c
    {$d\sigma/d\eta^{\ds} $ (nb)} \NL\RL
    &{$p_\perp^{\ds}>2\,\mbox{GeV}$}&{$p_\perp^{\ds}>3\,\mbox{GeV}$}\NL\RL
    $( -1.5, -1.0\,)$&$ 8.89 \E{0.81}{0.89}{0.33} $&
    $ 2.96 \E{0.23}{0.24}{0.26} $\NL
    $( -1.0, -0.5\,)$&$ 8.16 \E{0.80}{0.70}{0.48} $&
    $ 4.17 \E{0.29}{0.15}{0.26} $\NL
    $( -0.5,\+0.0\,)$&$ 7.61 \E{0.79}{0.97}{0.48} $&
    $ 3.88 \E{0.28}{0.25}{0.18} $\NL
    $(\+0.0,\+0.5\,)$&$ 5.23 \E{0.99}{1.09}{0.55} $&
    $ 2.93 \E{0.28}{0.18}{0.25} $\NL
    $(\+0.5,\+1.0\,)$&$ 3.21 \E{1.02}{0.86}{0.66} $&
    $ 2.11 \E{0.28}{0.19}{0.21} $\NL
    $(\+1.0,\+1.5\,)$&$ 4.65 \E{1.40}{1.49}{1.06} $&
    $ 2.32 \E{0.34}{0.46}{0.37} $\NL
    \RRL
    &{$p_\perp^{\ds}>4\,\mbox{GeV}$}&{$p_\perp^{\ds}>6\,\mbox{GeV}$}\NL\RL
    $( -1.5, -1.0\,)$&$ 1.021 \E{0.096}{0.079}{0.095} $&
    $ 0.052 \E{0.052}{0.048}{0.037} $\NL
    $( -1.0, -0.5\,)$&$ 1.641 \E{0.137}{0.074}{0.118} $&
    $ 0.331 \E{0.047}{0.054}{0.054} $\NL
    $( -0.5,\+0.0\,)$&$ 1.877 \E{0.143}{0.117}{0.112} $&
    $ 0.460 \E{0.061}{0.031}{0.039} $\NL
    $(\+0.0,\+0.5\,)$&$ 1.662 \E{0.137}{0.077}{0.067} $&
    $ 0.398 \E{0.056}{0.032}{0.028} $\NL
    $(\+0.5,\+1.0\,)$&$ 1.090 \E{0.117}{0.138}{0.079} $&
    $ 0.374 \E{0.045}{0.029}{0.036} $\NL
    $(\+1.0,\+1.5\,)$&$ 1.186 \E{0.142}{0.132}{0.137} $&
    $ 0.242 \E{0.062}{0.030}{0.043} $\NL
    \RRL
  \end{tabular}%
  \caption{The differential cross sections $d\sigma/d\eta^{\ds}$ for
    channel~(1) as function of $\eta^{\ds}$ for the kinematic regions
    of Fig.3.  The $\eta^{\ds}$ range is given in brackets. The quoted
    cross sections correspond to the centres of the corresponding
    bins. The first error is statistical and the second is systematic.
    Overall normalisation uncertainties due to luminosity measurement
    ($\pm 1.4\%$)and to $\ds$ and $\d0$ decay branching ratios ($\pm3.7 \%$)
    are not included in the systematic errors.}\smallskip
\end{table}

The results from the two $\d0$ decay modes are in good agreement and
are consistent with our published measurements based on data taken in
1994~\cite{dstar94}.

\subsection{Systematic Uncertainties}

A detailed study of possible sources of systematic uncertainties was
carried out for all the measured cross sections.  The numbers quoted
below are for the integrated cross section with
$p_\perp^{\ds}>2\,\mbox{GeV}$ of channel~(\ref{kpi}), unless stated
otherwise.

\begin{itemize}
\item Uncertainties originating from the modelling of the MC
  simulation were estimated from the difference in the cross sections
  obtained with the two event generators PYTHIA and HERWIG.  They are
  negligible for the cross section with $p_\perp^{\ds}>2\,\mbox{GeV}$;
  however they vary between $-2.2\%$ and $-4.5\%$ for the higher
  $p_{\perp}^{\ds}$ cuts.
\item To estimate the uncertainties in the tracking procedure, the
  track selection cuts were varied by $\pm 10\%$ from the nominal
  values (section~3.2).  The resulting combined uncertainty in the
  cross section is $^{+7.1}_{-2.2}\%$.  Changing the $p_{\perp}^{\ds}
  / E_{\perp}^{\theta > 10^{\circ}}$ cut by the same amount yields an
  uncertainty of $^{+0.8}_{-0.4}\%$.
\item The MC simulation was found to reproduce the absolute energy
  scale of the CAL to within $\pm 3\%$~\cite{F2}.  A shift of $\pm
  3\%$ due to the CAL energy scale uncertainty produces a variation of
  $^{+3.4}_{-2.6}\%$ in the cross section.  The dominant source of
  this uncertainty is due to the acceptance of the CAL energy
  thresholds in the FLT (section~2).  An additional uncertainty due to
  a small mismatch between data and MC in the observed CAL energy
  distribution amounts to $^{+1.5}_{-1.2}\%$.
\item Uncertainties in the background estimation of $^{+2.8}_{-0.4}\%$
  were obtained by varying the $\Delta M$ and $M(\d0)$ mass windows
  and the normalisation region (section~3.2).
\item The uncertainty from correcting $W_{JB}$ to the true $W$,
  determined by moving the $W_{JB}$ boundary values by the estimated
  resolution of $\pm 7\%$, was negligible for the cross section with
  $p_\perp^{\ds}>2\,\mbox{GeV}$.  For higher $p_{\perp}^{\ds}$ cuts
  the uncertainty varies between $-1.7\%$ and $+1.5\%$.
\item Reweighting the reference MC samples to other parton density
  parametrisations~\cite{pdf} for the proton (MRSA$^\prime$, GRV94HO,
  CTEQ3M) gave a variation of $^{+0.0}_{-1.5}\%$ in the cross section.
  Since the photon structure is not well known, we used several parton
  density parametrisations (LAC-G1, ACFGP, GS-G HO) and in addition we
  allowed a $\pm 10\%$ variation of the ratio of resolved to direct
  photon contributions with respect to the reference structure
  function. The largest resulting uncertainty in the cross section was
  $^{+4.1}_{-0.6}\%$.
\end{itemize}
All contributions to the systematic uncertainties, except the overall
scale uncertainties, were added in quadrature.  The combined
systematic uncertainties in the cross sections are given in Table~1.
For the differential cross sections the systematic errors were added
in quadrature to the statistical and are indicated in Figs.~2-4 by the
outer error bars.  In Tables~2 and 3 both types of errors are given
separately.

\section{Comparison with NLO QCD Calculations}

\subsection{ Massive Charm Scheme}

Full NLO calculations in the massive charm scheme of total and
differential cross sections for heavy quark production in the HERA
kinematic region have been published in ref.~\cite{NLOdiff}.  The
computation was done as in ref.~\cite{dstar94} for $\gamma p\to c\bar
c X$~\cite{NLO} and then converted to $e p\to c\bar c X$ cross section
with the appropriate flux factors~\cite{paper_93}.  The fraction of
$c$ quarks fragmenting into a $D^{*+}$ as measured by the OPAL
collaboration~\cite{OPAL}, $0.222\pm 0.014\pm 0.014$, was used to
produce total and differential $D^*$ cross sections in the restricted
kinematic regions of our measurements.

The calculation used the MRSG~\cite{MRSG} and GRV-G~HO~\cite{GRV}
parton density parametrisations for the proton and photon,
respectively.  The renormalisation scale used was \linebreak
$\mu_{R}=m_{\perp}=\sqrt{ m_{c}^{2} + p_{\perp}^{2}} $ ($m_{c} =
1.5$\,GeV) and the factorisation scales of the photon and proton
structure functions were set to $\mu_{F} = 2 m_{\perp}$.  The charm
fragmentation into $\ds$ was performed using the Peterson
function~\cite{PETER} $ f(z) \propto { \left[ z \left( 1 - { 1 / z } -
      { \epsilon / ( 1 - z ) } \right)^{2} \right] }^{-1}.$ Here $z$
is the fraction of the charm quark momentum taken by the $D^*$ and
$\epsilon$ is a free parameter.

The NLO cross sections obtained for the same kinematic regions as the
data are listed in Table~1 for $\epsilon$=0.02 and shown in Figs.~2
and~3 for $\epsilon$=0.02 (dashed lines) and $\epsilon$=0.06
(dash-dotted lines).  The value $\epsilon$=0.06 is based on
ref.~\cite{Chrin} and used by Frixione et al. in ref.~\cite{NLOdiff},
while $\epsilon$=0.02 is suggested by recent fits to $e^+ e^-$
data~\cite{cacciari}.  The predicted cross sections are considerably
lower than those measured.  The dotted line, which corresponds to the
extreme choice $\mu_R = 0.5 m_{\perp}$ and $m_c =1.2$~GeV, is still
below the data at the high $\eta^{\ds}$ regions.  The calculated
shapes in both the $p_{\perp}^{\ds}$ and $\eta^{\ds}$ distributions
are also inconsistent with the data.

The result of applying an effective intrinsic transverse momentum,
$k_T$, to the incoming partons in the massive charm scheme~\cite{NLO}
is relatively small. The predicted cross sections increase by about
$10\%$ with $\langle k_T^2 \rangle =1$~GeV$^2$, mostly at low $p_T$
and in the backward direction.  In a semi-hard approach~\cite{semi}
this effect was calculated according to the BFKL
evolution~\cite{BFKL}. Recently LO predictions using this approach
have become available~\cite{zotov}.  The predicted cross sections for
our kinematic range are close to the data in absolute value but do not
match the shape of the $\eta^{\ds}$ distribution.

\subsection{ Massless Charm Scheme}

A second type of NLO calculation~\cite{kniehl,kniehl2,cacciari}, the
massless charm scheme, assumes charm to be an active flavour in both
the proton and the photon.  The two massless charm calculations
factorise the perturbative and non-perturbative components of the
fragmentation differently and fit the latter part to the Peterson
function~\cite{PETER}, using recent $e^+ e^-$ data on $D^*$ production
to extract the $\epsilon$ parameter.  The fitted values obtained by
the two calculations in their specific factorisation schemes are
$\epsilon$=0.116~\cite{kniehl} and $\epsilon$=0.02~\cite{cacciari}.
Similar cross sections are obtained in each of the massless charm
calculations by fitting fragmentation functions other than the
Peterson one to the $e^+ e^-$ data.  These predictions are expected
not to be reliable when the minimum $p_{\perp}^{\ds}$ cut is as low as
2~GeV.

The cross sections predicted with these calculations~\cite{kniehl,
  cacciari} for the kinematic region of our measurement are listed in
Table~1 and shown as full lines in Figs.~2 and~3.  The parton density
parametrisations used were CTEQ4M~\cite{cteq} for the proton and
GRV-G~HO~\cite{GRV} for the photon.  The renormalisation and
factorisation scales as well as the values of $m_c$ are the same as in
the calculation of the massive charm approach.

The predictions of the two massless charm models give similar shapes
of the differential cross sections (Figs.2 and 3), but disagree with
each other in absolute magnitude by $\simeq$~40$\%$.  The cross
sections obtained by these predictions are mostly below the data.  In
particular the data are above the NLO expectations in the forward
direction.  The contribution of $D^*$ produced from $b\bar b$ in our
kinematic region, not included in the NLO curves, is
predicted\cite{kniehl2} to be below $5\%$, in agreement with our MC
estimation (section~4).  This fraction is found from the MC studies to
be slightly higher in the forward region, where it is up to $7\%$.

Using the MRSG~\cite{MRSG} parton density parametrisation of the
proton has no significant effect on the predictions.  In contrast, the
calculations depend on the parton density parametrisations of the
photon and in particular its charm content.  In order to check the
sensitivity of the $d\sigma /d\eta^{\ds}$ data to the parton density
parametrisation of the photon, we compare the results for
$p_{\perp}^{\ds} > 3$~GeV and $p_{\perp}^{\ds} > 4$~GeV in Fig.~4 with
the two NLO massless charm predictions~\cite{kniehl, cacciari}
obtained with the photon parton density parametrisations GRV-G
HO~\cite{GRV}, GS-G HO~\cite{GS} and AFG~\cite{AFG}.  The differences
between the various photon parton densities are at the $20\%$ level or
less in the integrated cross sections, but in the differential cross
sections considerable differences in shape are observed.  For the
massless charm scheme of ref.~\cite{kniehl}, the GS-G HO
curves~\cite{GS} are closest to the data.  However, in the GS-G HO
parton density function used for this calculation, charm and u-quarks
contribute equally.

\section{\boldmath Measurement of $D^*$ Dijet Cross Sections}

Given the discrepancies observed between data and NLO predictions in
the inclusive $\ds$ measurements, it is of interest to study the
kinematics of charm production in more detail. The measurement of jets
in the final state allows the kinematics of the hard scattering
process to be reconstructed.  In order to compare the measurement with
QCD calculations at any order, we define~\cite{xobs}
\begin{equation}
 \xgo = \frac{\Sigma_{\rm jets}(\ETJT\ e^{-\eta^{jet}})}{2 E_{e} y},
\end{equation}
where $\eta^{jet}$ is the jet pseudorapidity, $y$ is estimated by
$y_{JB}$, and the jets in the sum are the two highest $\ETJT$ jets
within the accepted $\ETAJ$ range.  The variable $\xgo$ is the
fraction of the photon momentum contributing to the production of the
two jets with the highest $\ETJT$.  In measurements, as well as in MC
simulations and higher order calculations, direct and resolved samples
can be separated by a cut on $\xgo$.  In this analysis we define a
direct (resolved) photon process by the selection $\xgo\ge 0.75
(<0.75)$.

Fig.\,5 shows the uncorrected transverse energy flow, $(1/N_{jet})dE_T
/d\Delta\eta$, around the jet axis (``jet profile") as a function of
$\Delta\eta =\eta^{CELL}-\eta^{jet}$, the distance in $\eta$ of the
CAL cell from the jet axis for the sample of dijet events associated
with a $D^*$ (section 3.3) with \linebreak $E_{T,cal}^{jet}>4$~GeV.
As for the inclusive $D^*$ analysis (section~3.2), wrong charge
combinations were used to subtract background from the $E_T$ flow in
the $D^*$ signal region. In order to reduce the uncertainties due to
the background subtraction procedure, a narrower $D^*$ region was used
in the jet profile plots: $1.82 < M(K\pi) < 1.90$~GeV and $0.144 <
\Delta M < 0.147$~GeV.  The jet sample is divided into three regions
of $\ETAJ$: $-2.4 < \ETAJ < 0.0$, $ 0.0 < \ETAJ < 1.0$ and $ 1.0 <
\ETAJ < 2.4$.  The distributions are plotted separately for direct
($\xgo\ge 0.75$) and resolved ($\xgo < 0.75$) events.  The jet
profiles are compared to the results of the HERWIG MC which includes
LO-direct and LO-resolved photon processes~\footnote{We distinguish
  between LO-direct and LO-resolved photon contributions using the LO
  diagrams as implemented in the MC simulation.}, shown as the full
histogram.  In inclusive dijet events~\cite{dijet}, the MC simulation
gives too little transverse energy in the forward (positive
$\Delta\eta$) region for low-$E_T$ jets, even when that simulation
includes MI. In contrast, our charm dijet $E_T$ flow distributions are
in reasonable accord with the MC without MI, including the forward
region.

Also shown in Fig.\,5 are the jet profiles obtained if only HERWIG
LO-direct photon events are used (dotted histogram). These profiles
have reduced $E_T$ flow in the backward (negative $\Delta\eta$) region
and do not describe the data with $\xgo < 0.75$, in particular for the
ranges $0 < \eta^{jet} < 1$ and $1 <\eta^{jet} < 2.4$.  The $E_T$ flow
in the backward direction is consistent with the presence of a remnant
from the resolved photon.

To calculate the cross section $d\sigma/d\xgo$ for dijets with an
associated $D^*$ meson, MC event samples have been used to correct the
charm dijet data for the efficiencies of the trigger and selection
cuts and for migrations caused by detector effects.  The resolution of
the kinematic variables was studied by comparing the MC simulated jets
reconstructed from final state particles (hadron jets) with jets
reconstructed from the energies measured in the calorimeter (detector
jets), and by comparing the corrected $y_{JB}$ with the true $y$. The
resolutions obtained are: in $\ETJT$ $\simeq 15\%$, in $\eta^{jet}$
$\simeq$ 0.1 and in $\xgo\simeq$ 0.06.  The correction factors are
calculated as the ratio $N_{\rm true}/N_{\rm rec}$ in each $\xgo$ bin,
where $N_{\rm true}$ is the number of events generated in a bin and
$N_{\rm rec} $ is the number of events reconstructed in that bin after
detector simulation and all experimental cuts.

Differential cross sections in $d\sigma/d\xgo$ in the range \wrang~,
$Q^2~<1$~GeV$^2$ are given for jets with $|\ETAJ| < 2.4$, $E_T^{jet1}
> 7$~GeV, $E_T^{jet2} > 6$~GeV and at least one $D^*$ in the range
$p_{\perp}^{\ds} > 3$~GeV, $-1.5 < \eta^{\ds} < 1.5$.  The asymmetric
cut on the hadron level $\ETJT$ values has been applied in order to
avoid a problem associated with a singularity in the NLO calculations
due to the soft gluons that accompany the jet~\cite{HO}.  The
increased minimum $p_{\perp}^{\ds}$ of 3~GeV compared to the inclusive
$D^*$ analysis (section~5) is due to the fact that there is almost no
$D^*$ signal in the region below this value due to the requirement of
the dijet cuts.  Background subtraction was performed as described in
section~3.2 for channel~(1).

\begin{table}[p]\tablesup
  \def\DR#1{\raisebox{1.75ex}[0pt][0pt]{$\left.
      \begin{array}{l}\null\\\null\end{array}\right\}\kern.5\tabcolsep #1$}}
  \smallskip
  \def\DR#1{\raisebox{1.75ex}[0pt][0pt]{$\left.
      \begin{array}{l}\null\\\null\end{array}\right\}\kern.5\tabcolsep #1$}}
  \begin{tabular}{ccc}\RRL
    $\xgo$ range& $d\sigma/d\xgo (nb)$ & $d\sigma/d\xgo (nb)$\NL\RL
    & $E_T^{jet1} > 7\,$GeV & $E_T^{jet1} > 6\,$GeV \NL
    & $E_T^{jet2} > 6\,$GeV & $E_T^{jet2} > 5\,$GeV \NL\RL
   (0.000--0.125)&
                  $0.32\E{0.19}{0.14}{0.26}\PM{0.00}{0.06}$&
                  $0.42\E{0.21}{0.33}{0.23}\PM{0.06}{0.08}$\NL
   (0.125--0.250)&
                  $1.06\E{0.30}{0.17}{0.22}\PM{0.10}{0.13}$&
                  $1.80\E{0.35}{0.53}{0.85}\PM{0.30}{0.20}$\NL
   (0.250--0.375)&
                  $1.20\E{0.28}{0.17}{0.36}\PM{0.17}{0.14}$&
                  $1.64\E{0.33}{0.52}{0.21}\PM{0.24}{0.17}$\NL
   (0.375--0.500)&
                  $0.98\E{0.31}{0.26}{0.26}\PM{0.15}{0.11}$&
                  $1.58\E{0.38}{0.33}{0.26}\PM{0.21}{0.15}$\NL
   (0.500--0.625)&
                  $1.24\E{0.27}{0.33}{0.24}\PM{0.18}{0.12}$&
                  $1.92\E{0.34}{0.52}{0.25}\PM{0.28}{0.17}$\NL
   (0.625--0.750)&
                  $1.80\E{0.36}{0.48}{0.20}\PM{0.24}{0.19}$&
                  $2.97\E{0.44}{0.30}{0.28}\PM{0.37}{0.32}$\NL
   (0.750--0.875)&
                  $3.70\E{0.53}{0.61}{0.65}\PM{0.54}{0.37}$&
                  $6.34\E{0.65}{0.62}{1.09}\PM{0.61}{0.58}$\NL
   (0.875--1.000)&
                  $2.87\E{0.37}{0.36}{0.33}\PM{0.23}{0.18}$&
                  $3.86\E{0.42}{0.45}{0.43}\PM{0.27}{0.16}$\NL
    \RL
   (0.000--0.250)&
                  $0.68\E{0.17}{0.12}{0.19}\PM{0.06}{0.09}$&
                  $ 1.10\E{0.20}{0.35}{0.48}\PM{0.26}{0.15}$\NL
   (0.250--0.500)&
                  $1.10\E{0.21}{0.17}{0.23}\PM{0.17}{0.14}$&
                  $ 1.63\E{0.25}{0.26}{0.16}\PM{0.24}{0.19}$\NL
   (0.500--0.750)&
                  $1.52\E{0.22}{0.30}{0.17}\PM{0.22}{0.16}$&
                  $ 2.43\E{0.27}{0.30}{0.16}\PM{0.34}{0.24}$\NL
   (0.750--1.000)&
                  $3.29\E{0.32}{0.42}{0.31}\PM{0.38}{0.35}$&
                  $ 5.10\E{0.38}{0.41}{0.62}\PM{0.44}{0.48}$\NL
    \RRL
  \end{tabular}
  \caption{The differential cross sections $d\sigma/d\xgo$ for
    channel~(1) as function of  $\xgo$ for the kinematic region 
    $E_T^{jet1} > 7$~GeV, $E_T^{jet2} > 6$~GeV, as given in Fig.6, and
    for the kinematic region $E_T^{jet1} > 6$~GeV, $E_T^{jet2} >
    5$~GeV. The $\xgo$ range is given in brackets. The quoted cross
    sections correspond to the centres of the corresponding bins.  The
    first error is statistical, the second is systematic and the third
    one is the energy scale uncertainty. Overall normalisation
    uncertainties due to luminosity measurement ($\pm 1.4\%$) and to
    $\ds$ and $\d0$ decay branching ratios ($\pm3.7 \%$) are not
    included in the systematic errors.}
\end{table}

The $d\sigma /d\xgo$ results are shown in Fig.6 and listed in Table~4.
All uncertainties except that due to the energy scale have been added
in quadrature.  The systematic uncertainty due to the energy scale is
shown in Fig.6 as the shaded band.  The cross section integrated over
$\xgo$ is {\xsecd}.  Results are also presented in Table~4 for the
region $E_T^{jet1} > $6~GeV, $E_T^{jet2} > $5~GeV, where the cross
section integrated over $\xgo$ is {\xsecc}.

\subsection{ Systematic Uncertainties}

Sources of systematic uncertainties in the cross section measurements
were investigated in a similar manner to section~5.1.  Additional
contributions specific to the $D^*$ and associated dijet sample for
the integrated cross sections in $\xgo$ with $E_T^{jet1} > $7~GeV,
$E_T^{jet2} > $6~GeV are:

\begin{itemize}
\item The possible shift in the CAL energy scale was increased to $\pm
  5\%$ due to the additional uncertainty in the $E_{T,cal}^{jet}$
  measurement~\cite{dijet}. The variation in the cross section is
  $^{+12.2}_{-9.8}\%$.
\item The uncertainty due to shifting the minimum $E_{T,cal}^{jet}$
  cut by $\pm 1$~GeV, which corresponds to the jet resolution in this
  low energy region is estimated to be $^{+2.1}_{-0.1}\%$.
\item Varying the $\eta^{jet}$ cut values by $\pm 0.1$ yields an
  uncertainty of $^{+0.1}_{-1.3}\%$.
\item Using the HERWIG MC with MI for the acceptance calculations
  contributes an uncertainty of $+1.1\%$.
\end{itemize}

All contributions to the systematic errors, excluding luminosity,
branching ratios and energy scale uncertainties, were added in
quadrature. The final systematic uncertainty in the total charm dijet
cross section is $^{+6.4}_{-3.9}\%$.  For the $\xgo$ differential
cross sections they were added in quadrature to the statistical errors
and are indicated as the outer error bars in Fig.\,6.  The energy
scale uncertainty is shown as the shaded bands.  Table~4 lists
separately the statistical, systematic and energy scale uncertainties.

\subsection{ Comparison with Theoretical Predictions }

In Fig.\,6(a) the $d\sigma /d\xgo$ distributions of the HERWIG MC
simulation, normalised to the data, are shown for the LO-direct and
LO-resolved contributions as well as for their sum.  The fractions of
each contribution was taken from the MC simulation.  There is a peak
in the data at high values of $\xgo$, consistent with a large
contribution from LO-direct photon processes.  However, there is also
a substantial tail to low $\xgo$ values, which is not described by the
LO-direct MC. Hence a LO-resolved component is required. In the
LO-resolved MC histogram, the dominant contribution from photon charm
excitation (lightly hatched) is distinguished from that of other
LO-resolved photon processes (densely hatched).  The contribution of
$b$-quarks to $D^*$ production was taken into account in the MC sample
as in the inclusive $D^*$ analysis.  It is about $10\%$ and
approximately constant with $\xgo$.  The MC distributions, where the
LO-resolved and LO-direct contributions are allowed to vary
independently, were fitted to the data. The data require a LO-resolved
contribution of $45 \pm 5 ~({\it stat.})\%$.  This value is consistent
with the LO HERWIG prediction of $37\%$. The charm excitation
contribution to the LO-resolved photon process in the HERWIG MC is
$93\%$.

A comparison of the data with a NLO calculation for a charm dijet
sample was performed using the massive charm approach~\cite{NLOdiff}.
This calculation does not have an explicit charm excitation component,
since charm is not treated as an active flavour in the photon
structure function.  The $\xgo$ distribution at the parton level was
estimated by applying the KTCLUS jet finder to the two or three
partons produced in this NLO calculation~\cite{NLO} for the kinematic
region of our $D^*$ and associated dijet analysis. Here $\epsilon =
0.02$ was used and $m_{\perp} = \sqrt{ m_{c}^{2} + \langle
  p_{\perp}^{2} \rangle } $, where $ \langle p_{\perp}^{2} \rangle$ is
the average $ p_{\perp}^{2} $ of the two charm quarks.  The result of
this calculation (full histogram) is compared to the data in Fig.\,6b.
To minimise migration effects due to hadronisation from high $\xgo$,
the data are given in wider bins compared to Fig.\,6a.  It can be seen
that the NLO massive charm calculation~\cite{NLOdiff} produces a tail
towards low $\xgo$ values similar to the light parton jet
case~\cite{jon}.  However, there is a significant excess in the data
over this NLO prediction. From MC studies we estimate that $\simeq
6\%$ of the highest $\xgo$ bin ($0.75 < \xgo < 1.0$) can migrate to
the lower bins due to hadronisation effects.  An effect of this size
cannot explain the measured low $\xgo$ cross section.  Using $\mu_R =
0.5 m_{\perp}$ and $m_c =1.2$~GeV in the calculation (dashed
histogram) yields a higher $\xgo$ tail, which is still below the data.
With these parameters the cross section near $\xgo$=1 is above the
data.  Applying an intrinsic transverse momentum $\langle k_T^2
\rangle =1$~GeV$^2$ (section~6.1) increases the predicted cross
sections in the two central $\xgo$ bins.  However the predicted cross
sections are still below the measurement.

The conclusions drawn above are the same when: a) the hadron level jet
cuts $E_T^{jet1} > $6~GeV, $E_T^{jet2} > $5~GeV (Table~4) were used;
b) a cone jet algorithm~\cite{cone} was applied instead of a cluster
algorithm; c) PYTHIA MC was used instead of HERWIG; d) the jet trigger
described in section 2 was used instead of the nominal one.

The extent to which a NLO calculation of charm in association with
dijets may describe the $d\sigma /d\xgo$ distribution must await
further theoretical developments. In particular additional
contributions arising from a photon structure function require
massless charm NLO predictions for $d\sigma /d\xgo$.

\section{Summary and Conclusions}

The integrated and differential inclusive photoproduced $\dspm$ cross
sections in $ep$ collisions at HERA have been measured with the ZEUS
detector in the kinematic region $Q^2 < 1$\,GeV$^2$, \wrang,
\ptrang~and \etarang. The cross section
$\sigma_{ep\,\rightarrow\,\dspm\,X}$={\xseca} was measured using the
channel~$\dskpi$.  A second $\ds$ decay channel has been studied,
$\dsk3pi$, and good agreement with the $K \pi$ channel has been found
in the region of overlap ($p_{\perp}^{\ds} > 4$\,GeV).  The results
are compared with massive and massless charm scheme QCD NLO
predictions.  The NLO calculations are generally below the measured
cross sections, in particular in the forward direction.  The results
are sensitive to the parton density parametrisation of the photon used
to calculate the cross section in the massless charm scheme.

A sample of inclusive dijet events with an associated $D^*$ meson has
been used to measure the cross section $d\sigma/d\xgo$ in the range
\wrang~ and $Q^2 < 1$\,GeV$^2$. The jets were reconstructed with the
KTCLUS algorithm, requiring $|\ETAJ| < 2.4$ and at least one $D^*$ in
the range \etarang~and \ptrangold .  Cross sections are given for the
kinematical regions $E_T^{jet1} > 7$~GeV, $E_T^{jet2} > 6$~GeV and
$E_T^{jet1} > 6$~GeV, $E_T^{jet2} > 5$~GeV.  A peak at high values of
$\xgo$ is seen, in agreement with the expectation for direct photon
processes. A large cross section is also measured at low $\xgo$, where
resolved processes are expected to contribute significantly.  A
comparison of the $\xgo$ distribution to MC simulations yields a
contribution to the cross section of about $45\%$ from LO-resolved
photon processes and indicates the existence of charm excitation in
the photon parton density. The data at $\xgo < 0.75$ are higher than a
NLO massive charm calculation at the parton level.

\section{Acknowledgements}

We would like to thank the DESY Directorate for their strong support
and encouragement.  The remarkable achievements of the HERA machine
group were essential for the successful completion of this work and
are greatly appreciated. We would like to thank M. Cacciari, S.
Frixione and B. Kniehl for discussions and for providing their NLO
calculations.

\clearpage
\begin{figure}[ht]
\begin{center}
\noindent\epsfig{file=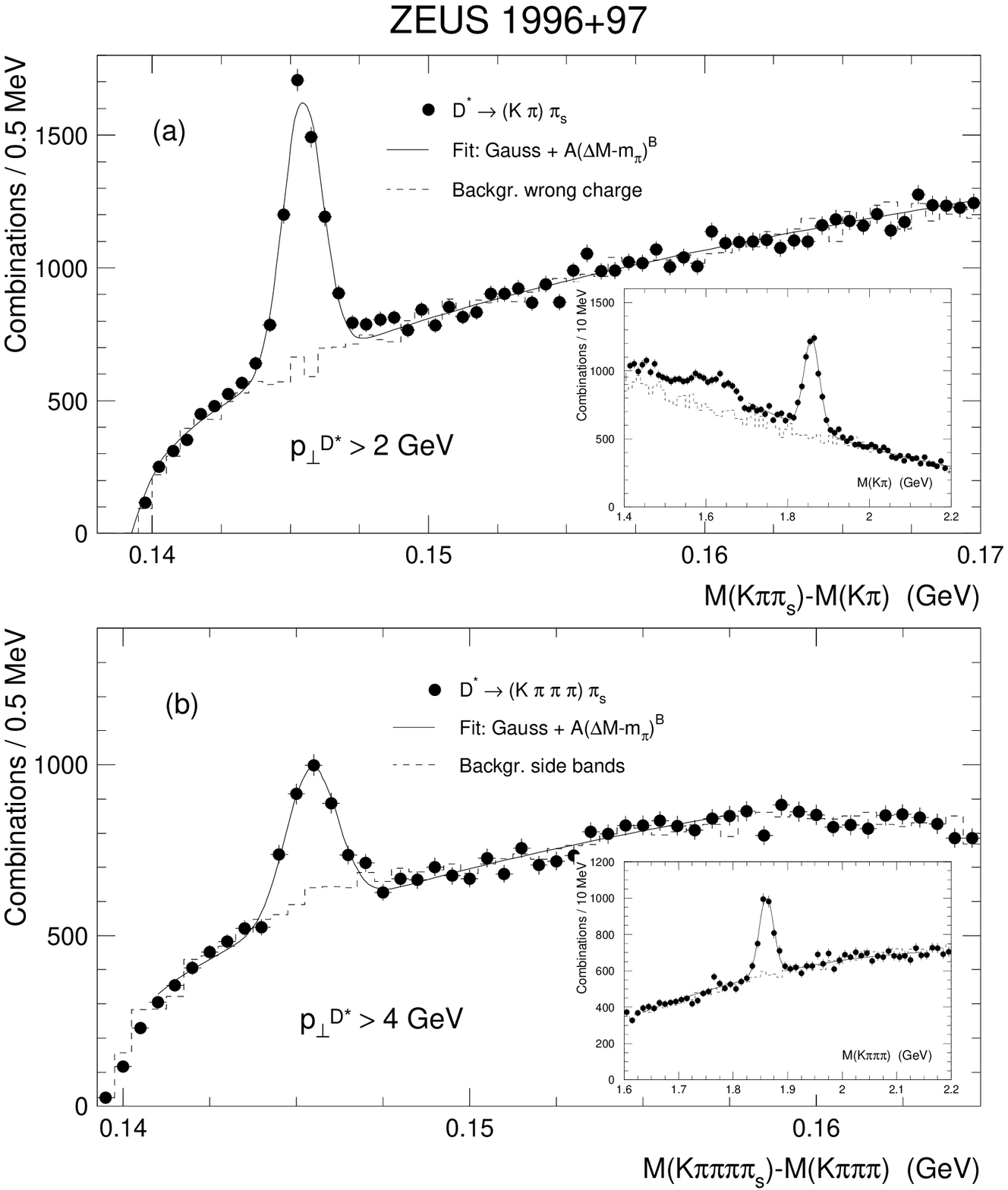, width=0.95\hsize}
\caption{$\Delta M$ distributions (a) for the $(K\pi)\pi_S$ channel with
  $p_{\perp}^{\ds} > 2$~GeV~ and (b) for the $(K\pi\pi\pi)\pi_S$
  channel with $p_{\perp}^{\ds} > 4$~GeV.  The full dots are right
  charge combinations from the $\d0$ signal region ($1.80 -
  1.92$~GeV). The dashed histograms are wrong charge combinations from
  the $\d0$ region for the $(K\pi)\pi_S$ channel and side bands
  combinations (see text) for the $(K\pi\pi\pi)\pi_S$ channel.  The
  full lines are the results of fits to a sum of a Gaussian and the
  functional form $A\cdot (\Delta M-m_{\pi})^{B}$.  The insets in (a)
  and (b) are the $M(K\pi)$ and $M(K\pi\pi\pi)$ distributions from
  combinations having $143 < \Delta M < 148$~MeV.  The dashed
  histograms are wrong charge and side bands combinations,
  respectively.  }
\end{center}
\end{figure}

\clearpage
\begin{figure}[b]
\begin{center}
\noindent\epsfig{file=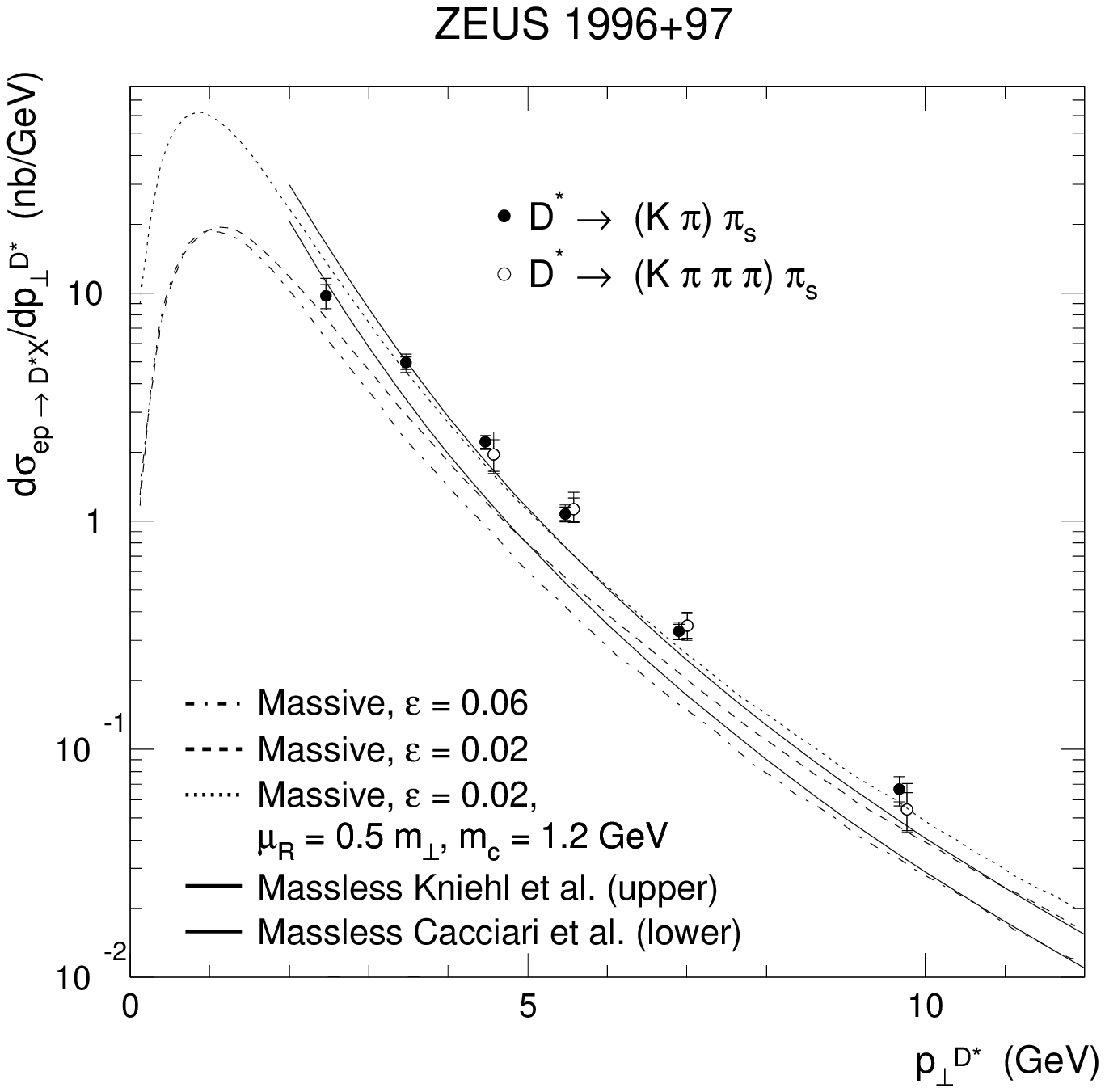, height=13.5cm}
\caption{The differential cross section $d\sigma/dp_{\perp}^{\ds}$ for
  $\ds$ photoproduction, \qqrang, in the kinematic region \wrang~and
  \etarang~for the $(K\pi)\pi_S$ (full dots) and $(K\pi\pi\pi)\pi_S$
  (open dots) channels.  The $(K\pi)\pi_S$ points are drawn at the
  positions of the average values of an exponential fit in each bin.
  The $(K\pi\pi\pi)\pi_S$ points are offset for clarity.  The inner
  part of the error bars shows the statistical error, while the outer
  one shows the statistical and systematic errors added in quadrature.
  The predictions of NLO perturbative QCD calculations are given by
  the dash-dotted, dashed and dotted curves for the massive charm
  approach~\cite{NLOdiff} and by the full upper (lower) curve for the
  massless charm approach calculation of ref.~\cite{kniehl}
  (ref.~\cite{cacciari}), with the parameters described in section~6.
  }
\end{center}
\end{figure}

\clearpage
\begin{figure}[b]
\begin{center}
\noindent\epsfig{file=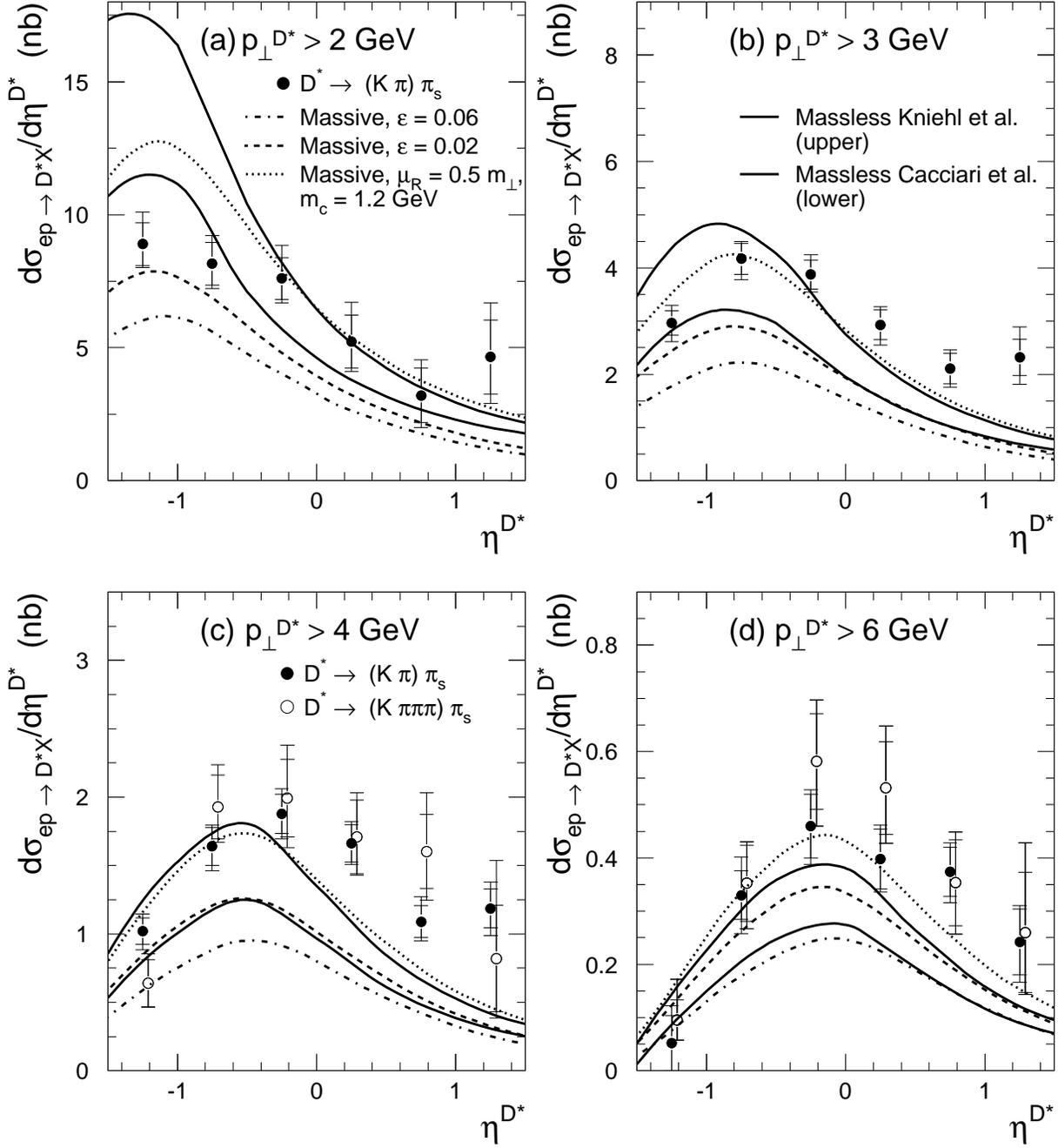, width=\hsize}
\caption{ Differential cross sections $d\sigma/d\eta^{\ds}$ for $D^*$
  photoproduction, \qqrang~, in the kinematic region \wrang~and
  a) $p_{\perp}^{\ds} > 2$~GeV;
  b) $p_{\perp}^{\ds} > 3$~GeV;
  c) $p_{\perp}^{\ds} > 4$~GeV;
  d) $p_{\perp}^{\ds} > 6$~GeV.  
  The $(K\pi)\pi_S$ points are drawn at the centres of the
  corresponding bins.  The $(K\pi\pi\pi)\pi_S$ points are offset for
  clarity.  The inner part of the error bars shows the statistical
  error, while the outer one shows the statistical and systematic
  errors added in quadrature.  The curves correspond to the same
  predictions of NLO perturbative QCD calculations as in Fig. 2.}
\end{center}
\end{figure}

\clearpage
\begin{figure}[b]
\begin{center}
\noindent\epsfig{file=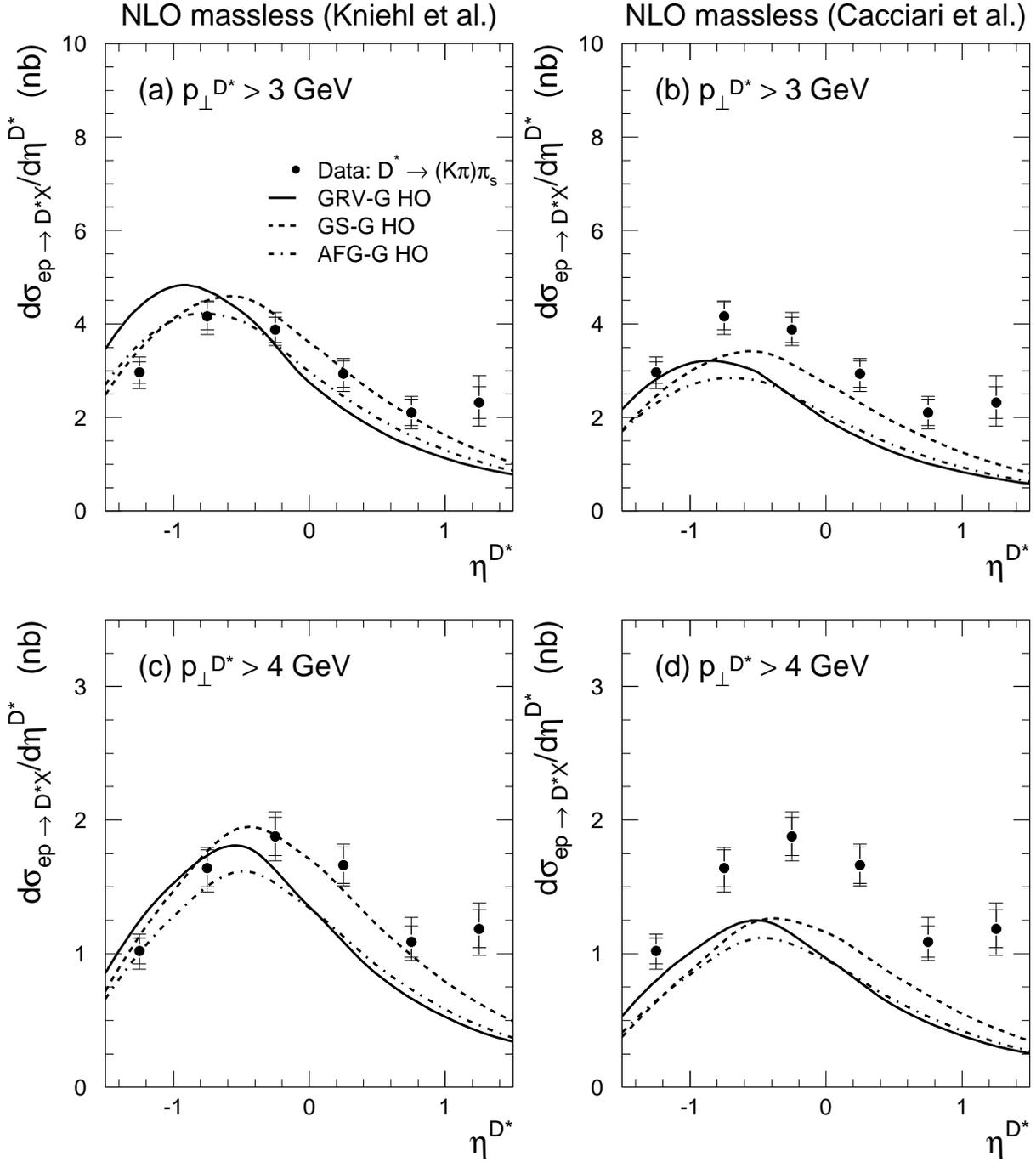, width=\hsize}
\caption{ Differential cross sections $d\sigma/d\eta^{\ds}$ for $D^*$
  photoproduction, \qqrang~, in the kinematic region \wrang~for the
  $(K\pi)\pi_S$ channel with
  (a-b)$p_{\perp}^{\ds} > 3$~GeV and
  (c-d)$p_{\perp}^{\ds} > 4$~GeV.
  The points are drawn at the centres of the corresponding bins.  The
  inner part of the error bars shows the statistical error, while the
  outer one shows the statistical and systematic errors added in
  quadrature.  The curves are the predictions of the massless charm
  NLO of ref.~\cite{kniehl} (a,c) and ref.~\cite{cacciari} (b,d) with
  various photon parton density parametrisations.}
\end{center}
\end{figure}

\clearpage
\begin{figure}[b]
\begin{center}
\noindent\epsfig{file=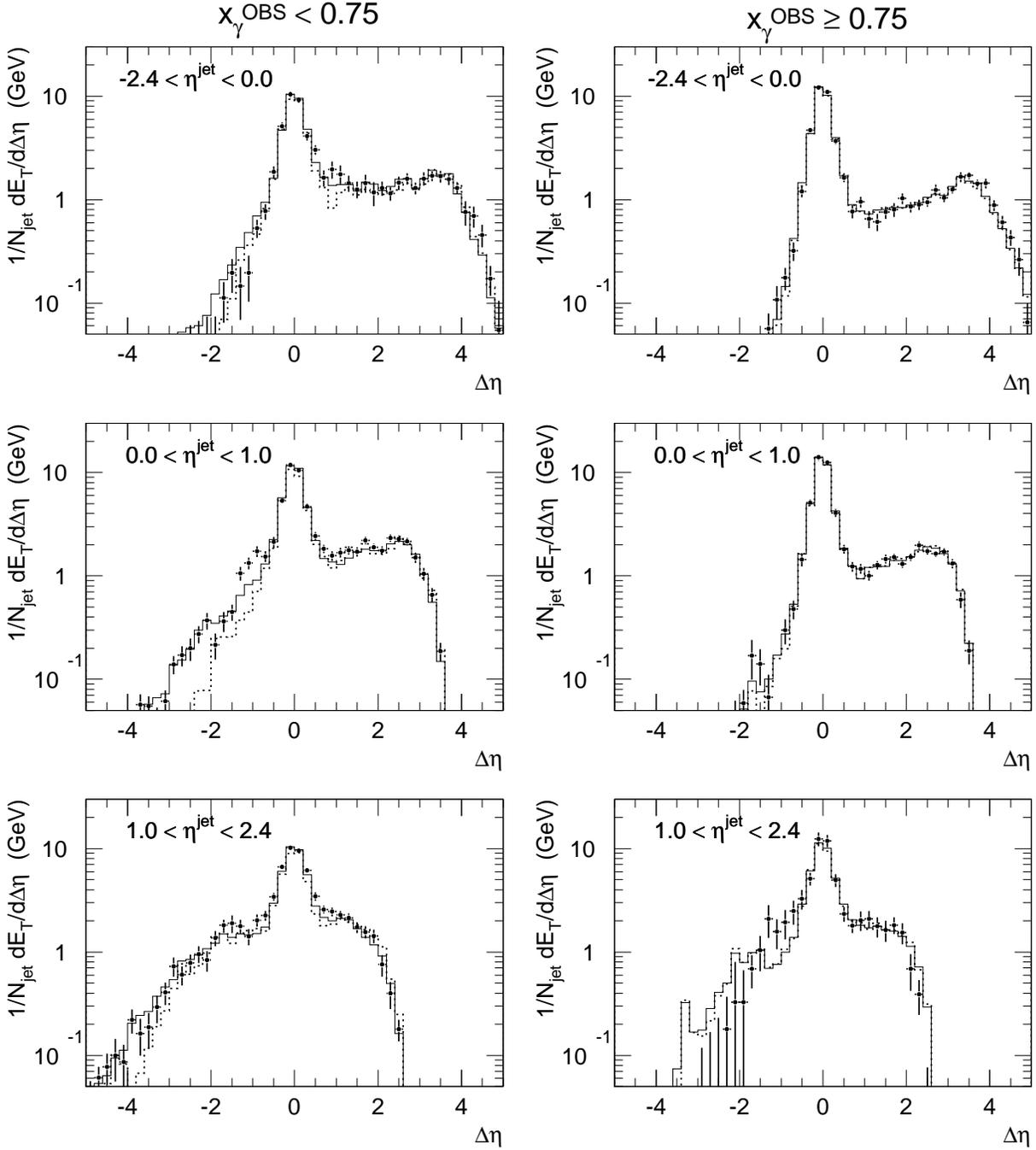, width=\hsize}
\caption{Uncorrected transverse energy flow with respect to the jet axis for
  dijet events containing a $D^*$ in a kinematic region given in the
  text and for $E_{T,cal}^{jet} > 4$~GeV.  The jets are defined using
  the KTCLUS jet algorithm. The distributions are given in three
  regions of $\eta^{jet}$ separately for direct ($\xgo\ge 0.75$) and
  resolved ($\xgo < 0.75$) photon events.  The data (dots) are
  compared to expectations of the HERWIG MC (full histogram) and
  LO-direct only (dotted histogram).  The error bars represent the
  statistical uncertainty only.}\label{f:profiles}
\end{center}
\end{figure}

\clearpage
\begin{figure}[b]
\begin{center}
\noindent\epsfig{file=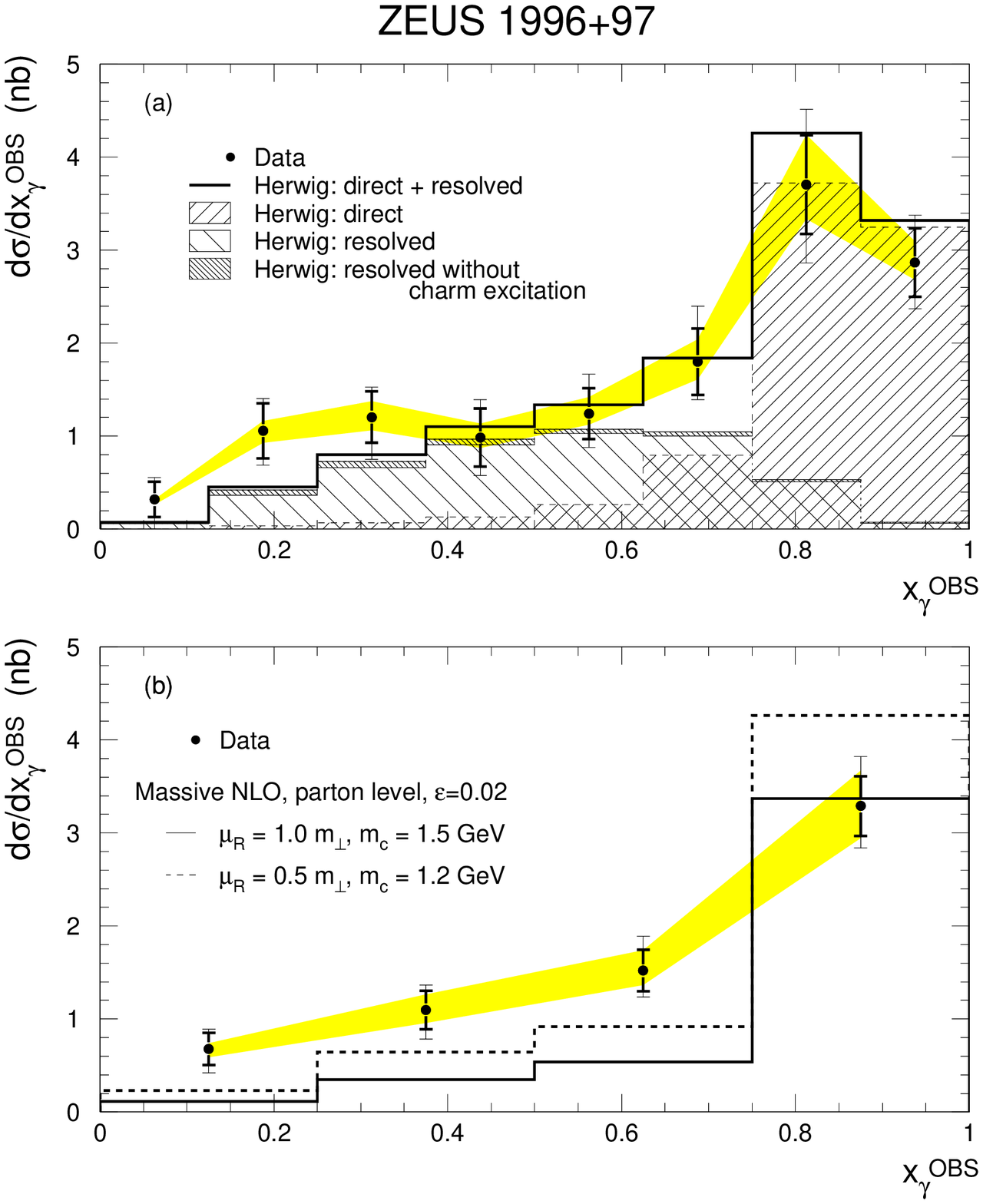, width=.82\hsize}
\caption{ The differential cross section $d\sigma/d\xgo$ for dijets
  with an associated $D^*$ meson with $p_{\perp}^{\ds} > 3$~GeV, $-1.5
  < \eta^{\ds} < 1.5$ in the kinematic range \wrang, $Q^2 <
  1$~GeV$^2$, $|\ETAJ| < 2.4$, $E_T^{jet1} > 7$~GeV and $E_T^{jet2} >
  6$~GeV.  The KTCLUS algorithm is used for the jet definition.  The
  points are drawn at the centres of the corresponding bins.  The
  inner part of the error bars shows the statistical uncertainties.
  The outer part is the statistical and systematic errors added in
  quadrature.  The energy scale uncertainty is given separately by the
  shaded bands.  In (a) the experimental data (dots) are compared to
  the expectations of the HERWIG simulation, normalised to the data,
  for LO-direct (right hatched), LO-resolved (left hatched),
  LO-resolved without charm excitation (dense hatched) and the sum of
  LO-direct and LO-resolved photon contribution (full histogram).  In
  (b) the data are compared with a parton level NLO massive charm
  calculation~\cite{NLOdiff} with the parameters described in
  section~7.2.
\label{f:xgo}}
\end{center}
\end{figure}
\end{document}